# The prediction of meteor showers from all potential parent comets

**Lubos Neslusan[1], Maria Hajdukova[1],**
**Dusan Tomko[1], Zuzana Kanuchova[1] and Marian Jakubik[1]**

[1] Astronomical Institute, Slovak Academy of Sciences, Slovakia
`Maria.Hajdukova@savba.sk`
`[ne;kanuchova;dtomko;mjakubik@ta3.sk`

The objectives of this project are to predict new meteor showers associated with as many as possible known periodic comets and to find a generic relationship of some already known showers with these comets. For a potential parent comet, we model a theoretical stream at the moment of its perihelion passage in a far past, and follow its dynamical evolution until the present. Subsequently, we analyze the orbital characteristics of the parts of the stream that approach the Earth's orbit. Modelled orbits of the stream particles are compared with the orbits of actual photographic, video, and radar meteors from several catalogues. The whole procedure is repeated for several past perihelion passages of the parent comet. To keep our description compact but detailed, we usually present only either a single or a few parent comets with their associated showers in one paper. Here, an overview of the results from the modelling of the meteor-shower complexes of more than ten parent bodies will be presented. This enables their diversities to be shown. Some parent bodies may associate meteor showers which exhibit a symmetry of their radiant areas with respect to the ecliptic (ecliptical, toroidal, or showers of an ecliptic-toroidal structure), and there are showers which have no counterpart with a similar ecliptical longitude on the opposite hemisphere. However, symmetry of the radiant areas of the pair filaments with respect to the Earth's apex is visible in almost all the complexes which we examined.

## 1 Introduction

The aim of this project is to reveal alterations in initial orbital corridors of meteoroid streams which were formed due to gravitational action. This enables generic relationships between a meteoroid stream and a parent comet that do not have similar orbits to be found and new meteor showers to be predicted. Stream meteoroids which move each in an orbit similar to the comet's orbit create around the comet's orbit a spatial orbital corridor. If the stream is not significantly perturbed, the orbit of the parent comet will be situated inside the corridor at its center.

If this orbit is situated at a large distance from the orbit of our planet, the particles of the stream do not usually collide with its atmosphere and thus create a meteor shower. However, in some cases, the gravitational perturbations of major planets can deflect a significant number of particles from the corridor around the parent body orbit, being far from the Earth's orbit, into an alternative corridor crossing this orbit. Thus, some cometary or asteroidal objects in distant orbits can still have associated a stream colliding with the Earth's atmosphere.

If the entire orbit of a comet is situated relatively close to the orbit of the Earth, the particles also pass relatively close to the orbit of our planet, and some of them collide with its atmosphere. Here also, the perturbations can change the orbits of a part of a stream. As a consequence, an alternative corridor, or corridors, of orbits are formed. If more than a single corridor of a given stream passes through the Earth's orbit, we observe several meteor showers associated with the same parent body.

All these alterations of the initial orbital corridors can be revealed by our modelling theoretical streams and studying their dynamical evolutions for a suitably long period. So far, we have modelled meteor-shower complexes of eleven parent bodies, the analyses of which have already been individually published in the following papers: Neslušan (1999), Kaňuchová & Neslušan (2007), Tomko & Neslušan (2012), Neslušan et al. (2013a, b), Tomko (2014), Neslušan & Hajduková (2014), Neslušan et al. (2014), Tomko & Neslušan (2014). Here, we present an overview of these results, comparison of which enables their diversities and/or similarities to be shown. The procedure allows us to map the whole complex of meteoroid particles released from a parent comet. The relationship between a particular comet and known showers can be either confirmed or shown doubtful, or a new relationship can appear. The structure of modeled complexes demonstrates the distribution of the cardinal directions of meteor sources and contributes to the map of the whole meteoroid population in the Solar System.

## 2 Modeling theoretical streams

Meteoroid streams of several parent bodies and their dynamical evolution were studied with the help of various stream models (e.g., Williams & Wu 1994; Brown & Jones 1998; Asher 1999; Beech 2001; Asher & Emel'yanenko 2002; Lyytinen & Jenniskens 2003; Jenniskens 2004; Williams et al. 2004; Asher 2005; Vaubaillon et al. 2005a,b; Wiegert et al. 2005; Wiegert & Brown 2005; Porubcan & Kornos 2005; Vaubaillon & Jenniskens 2007; Asher 2008; Babadzhanov et al. 2008; Wiegert et al. 2009; Jenniskens & Vaubaillon 2010; Babadzhanov et al. 2013; Jopek et al. 2013; Sekhar & Asher 2014 a,b).

We decided to base our study of the dynamics of meteoroid streams on the gravitational action, exclusively. Since we aim to predict new streams, we do not consider any non-gravitational effects. We assume that including these effects, characterized by the entire ranges of possible free parameters, would only enlarge the dispersion of predicted stream characteristics. Our aim is not to model the stream in all its details corresponding to a realistic scenario (which is never completely known in a specific case), but to cover the orbital phase space of the densest core of an actual stream only. The modelled set of orbits represents the most central part of the stream, not the entire stream.

For a potential parent comet, we model a theoretical stream at the moment of its perihelion passage in a distant past, monitor its orbital evolution up to the present, and select a part of the stream that approaches the Earth's orbit. These particles are used to predict the corresponding meteor showers. The predicted showers are searched for in the databases of actually observed meteors. The whole procedure is repeated for several past perihelion passages of the parent body. The procedure allows us to map the whole complex of meteoroid particles released from a parent comet. Detailed description of the method can be found in the individual papers mentioned in the Introduction.

## 3　Analyses and results

Our modelling has revealed new relationships among the known meteor showers observed in the Earth's atmosphere that belong to the same complex. New parent bodies associated with known meteor showers have been found, and new meteor showers have been predicted to be observed. The results of our project are summarized and briefly listed in the following paragraphs. All the results contributed significantly to the task of finding parent bodies of minor meteor showers, which is one of the subjects of recent meteor research (Wiegert & Brown 2004; Brown et al. 2010; Rudawska et al. 2012, Rudawska & Vaubaillon 2014).

**Overlapping meteor-shower complexes of 14P/Wolf and D/1892 T1 (Barnard 3)**

The dynamical study of the meteor stream associated with comet 14P/Wolf shows that the planetary gravitational disturbances split the corridor of the stream into several filaments. Meteors of two of them can enter the Earth's atmosphere and become observable. The model corresponds to the orbit of the comet 14P/Wolf before 1922, when the comet was moved to a new orbit by the gravitational disturbance of Jupiter and stopped releasing meteoroid particles into the orbits crossing the orbit of the Earth.

The filament with a higher declination of mean radiant coincides with the meteor stream associated with comet D/1892 T1 (Barnard 3). The filament of 14P/Wolf stream with a lower declination of radiant coincides with the well-known meteor shower α-Capricornids (Neslušan, 1999). We note that Jenniskens and Vaubaillon (2010) found a better parent body of the α-Capricornid shower: the minor planet 2002 EX12 (=169P/NEAT). A possible dynamical relationship of 14P with this asteroid has not been investigated yet.

**Nearly identical meteor-shower complexes of the comet 96P/Machholz and asteroid 2003 EH1**

Theoretical streams of the comet 96P/Machholz and asteroid 2003 EH1 evolved, after a significant time, into almost identical structure Neslušan et al. (2013a, b), Kaňuchová & Neslušan (2007). Both streams are highly structured. The 96P approaches the Earth's orbit in six intervals of the ecliptic longitude of Earth. (There are eight approaches in total, but two of these intervals partially overlap.) The particles in three of these intervals hit the Earth from the northern direction and in the other three from the southern direction relative to the ecliptic. As a consequence, we can distinguish six filaments (F1-F6) in the part of the stream that approaches the orbit of Earth. These filaments correspond to six meteor showers with the radiants distributed on the sky symmetrically with respect to the Earth's apex. Four of them are well known from observations: daytime Arietids (F1), δ-Aquarids S (F5) and N (F2), and Quadrantids (F3). Filament 4, predicted to be the southern branch of the daytime Arietids, could be, with an uncertainty, identified with α-Cetids. Its identification, as well as that of filament 6 (with similar characteristics as the κ-Velids, or the Puppid-Velid Complex, or the Carinid Complex), with corresponding showers in the considered databases were negative. The complex structure of the stream is demonstrated with the predicted radiants of particles expected to collide with the Earth in the present. The positions of the radiants are plotted in Fig. 1 left.

Similarly, examination of the asteroid 196256 (2003EH1) showed that six well-established and two minor filaments approach the orbit of the Earth, producing the same four well-known meteor showers as the stream 96P (Fig 1 right). If we followed a longer evolutionary period, then another shower having both northern and southern branches occurred (filaments 7 and 8). If this shower exists, it would be an ecliptical shower related to the Arietids.

**Meteor-shower complex of the long-period comet C/1917 F1 (Mellish)**

We also focused our attention on the meteor-shower complex of the comet C/1917 F1 Mellish (Neslušan & Hajduková 2014). The modeled complex is shown in Fig. 2. The theoretical stream split into four filaments. We have confirmed the generic relationship between the studied parent comet and the December Monocerotids (F3). The comet is probably also the parent body of the April ρ-Cygnids (F1). The evolution of meteoroids to the orbits of this shower is very long at about 20 millennia. Following even a longer evolutionary period, up to 50 millennia, two other diffuse showers with the radiants situated symmetrically to both the December Monocerotids and April ρ-Cygnids showers with respect to the apex of the Earth's motion occurred (filaments 2 and 4). However, we did not find any corresponding shower in the list of the IAU.

Our simulation did not confirm any relationship between C/1917 F1 and the November Orionids, a shower which is, according to several authors, related to the comet Mellish (Veres et al. 2011).

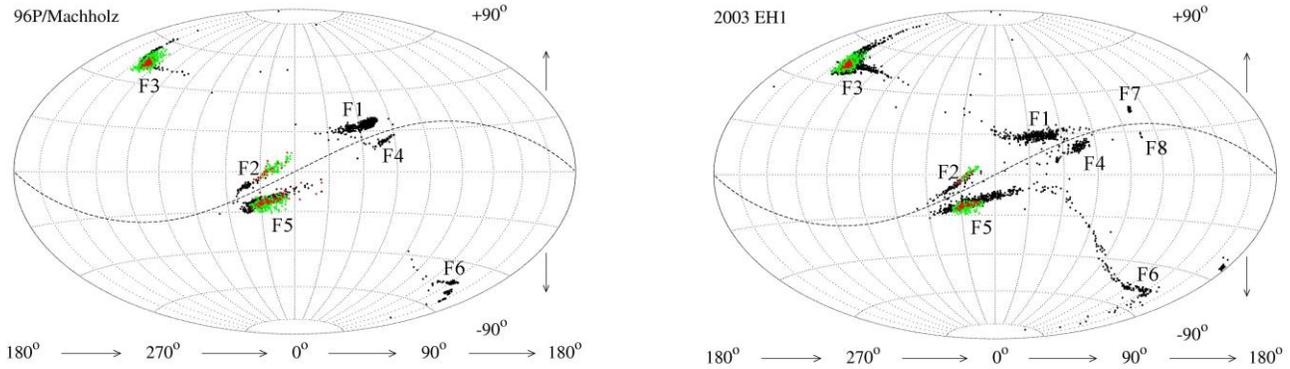

*Figure 1*-Nearly identical meteor-shower complexes of the comet 96P/Machholz and the asteroid 2003 EH1. The radiants calculated from the modeled orbits (black dots) are compared with those of the real meteors from the video (green) and photographic (red) catalogues (SonotaCo 2009; Lindblad et al. 2003). Identified associated showers: daytime Arietids (F1), northern (F2) and southern (F5) branch of δ-Aquarids, Quadrantids (F3); possibly associated showers: α-Cetids (F4), κ-Velids, or the Puppid-Velid Complex, or the Carinid Complex (F6). Filaments F7 and F8, which occurred in the model of the asteroid, could not be identified to any showers of the IAU MDC list. Positions of the radiants in right ascension and declination are shown in the Hammer projection of equatorial coordinates. The sinusoid-like curve illustrates the ecliptic.

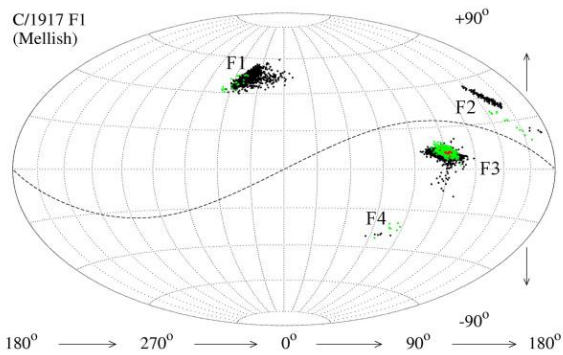

*Figure2*-Radiants of the modeled meteor-shower complex of the comet F1/1917 (Mellish) compared with those of the real meteors." Identified showers: December Monocerotids (F3) and possibly April ρ Cygnids (F1). We did not find any corresponding showers in the IAU MDC list to filaments F2 and F4.

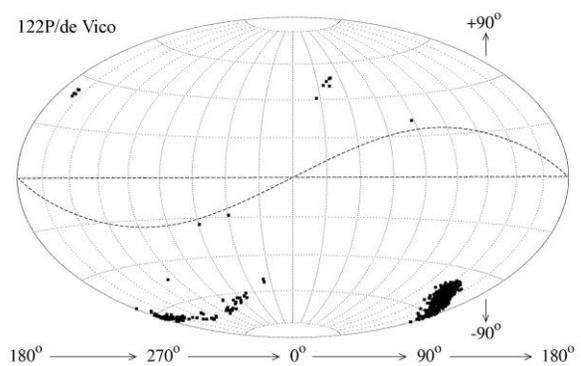

*Figure 3*-Radiants of new meteor-shower, associated with the comet 122P/de Vico, which is predicted in the southern hemisphere. Identification with real meteors was negative, probably because a low number of real meteors in the southern sky have been detected.

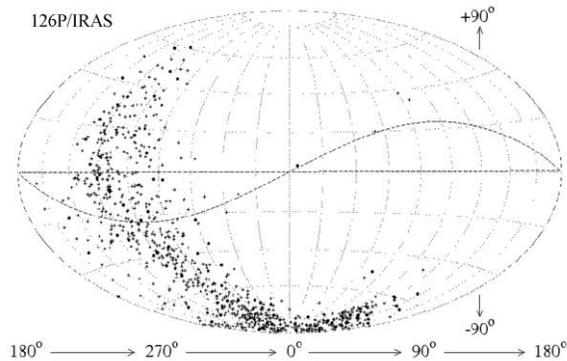

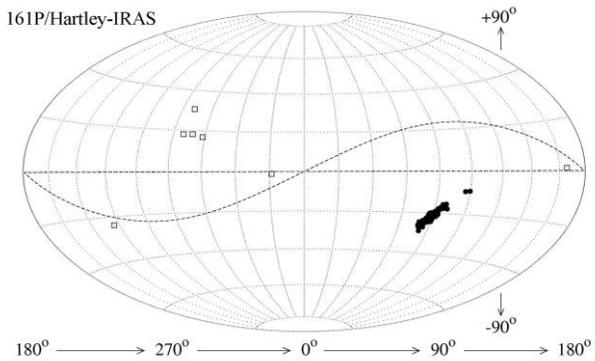

*Figure 4*-Parts of theoretical streams of two comets in orbits situated at a relatively large distance from the orbit of Earth that cross, according to our modeling, the Earth's orbit and, eventually, could be observed as meteors. However, the radiant area of meteors of the 126P/2004 V2 (left) is largely dispersed and, therefore, mixed with the sporadic meteor background. On the other hand, there seems to be a quite high chance of discovering the shower of 161P/2004 V2 (right) with a compact radiant area on the southern sky.

**New meteor showers in the southern hemisphere predicted**

We modelled theoretical streams of several comets, associated meteor showers of which have predicted radiant areas on the southern sky. The identification with real meteors is, in these cases, difficult because southern sky observations are significantly rarer than observations of the northern sky. Thus, a low number of real meteors of the southern sky has been detected and, therefore, recorded in the databases used.

Two Halley-type comets, 161P/2004 V2 and 122P/de Vico, that were examined, associate meteor showers with compact radiant areas on the southern sky (Fig. 3, 4 right). We point out that, in spite of a relatively large distance of the orbit of 161P/2004 V2 from the Earth (the minimum distance between the 161P orbit and the Earth's orbit was largest, about 1.8 AU, 4300 years before the present, and it does not, at the present, approach the Earth's orbit closer than 0.4 AU), the comet may associate an Earth-observable meteor shower.

A significant fraction of particles released from another studied comet in a distant orbit, 126P/2004 V2, also cross, according to our modelling, the Earth's orbit and, eventually, could be observed as meteors, prevailingly on the southern sky. However, their radiant area is largely dispersed (declination of radiants spans from about +60 to the southern pole) and, therefore, mixed with the sporadic meteor background (Fig. 4 left). Identification with real meteors is practically impossible.

The question of the existence of these showers remains open. However, there seems to be quite a high chance of discovering at least some of them in the future with an expected increase of observations of the southern hemisphere. More detailed description is in the papers Tomko & Neslušan (2012), Tomko (2014), Tomko & Neslušan (2014).

## 4   Symmetry on the sky

An interesting issue concerns the symmetry of filaments of the modelled stream. The radiants of showers of all complexes which consist of more than one filament are distributed on the sky symmetrically with respect to the Earth's apex. Some of them are symmetrical to the ecliptic. Some of them have no counterpart with a similar ecliptical longitude on the opposite hemisphere. But, still, they exhibit a symmetry with respect to the Earth's apex.

The cardinal directions of meteor sources, "helion","antihelion" and "northern and southern toroidal" are more often related to the sporadic meteor background since the ecliptic-toroidal structure appears in the overall distribution of radio-meteor radiants, as published by Campbell-Brown & Brown (2005). According to Jenniskens (2006), many well-known streams, such as Daytime Arietids, δ-Aquarids and others belong to the ecliptical streams in the sense of cardinal meteor directions. Obviously, there are several parent bodies feeding these streams. Our study has showed a relationship between ecliptical streams and toroidal streams. The meteor-shower complexes of comet 96P and asteroid 2003 EH1 showed that a single parent body can associate showers of both kinds, ecliptical and toroidal (Neslušan et al. 2013a, b, 2014). The ecliptic-toroidal structure is seen transparently in these models (Fig. 5 b, d). The predicted radiants of particles that approach Earth's orbit are plotted in the ecliptical coordinate system modified to place its origin at the apex of the Earth's motion. The filaments corresponding to the Arietids, δ-Aquarids S and N, and possibly α-Cetids constitute the ecliptical component and those corresponding to the Quadrantids, and possibly κ-Velids, constitute the toroidal component of the complex.

Meteor showers of the long-period comet C/1917 F1 Mellish (Fig. 5a) have no counterpart with a similar ecliptical longitude on the opposite hemisphere, but they show a certain symmetry of their showers' radiant areas with respect to the Earth's apex (Neslušan & Hajduková 2014).

## 5   Conclusion

Within the frame of the project "The prediction of meteor showers from all potential parent comets", we have so far investigated 11 parent bodies. Our procedure is based on the modelling of a theoretical stream for several moments of the perihelion passages of a parent body in the distant past, monitoring its orbital evolution up to the present, selecting that part of the stream which approached the Earth's orbit, and comparing the characteristics of this part with the corresponding observed meteor shower. New meteor showers, mainly in the southern hemisphere, were predicted and new parent bodies of meteor showers, resp. new relationships between observed showers were suggested.

A comparison of the theoretical streams of several examined comets enabled their diversities and specific features to be shown. The results are summarized in the following conclusions:

- a single parent body can associate multiple showers

- a shower can be associated to multiple parent bodies

- shower radiants of a complex are distributed on the sky symmetrically with respect to the Earth's apex

- an ecliptic-toroidal structure of complexes was found

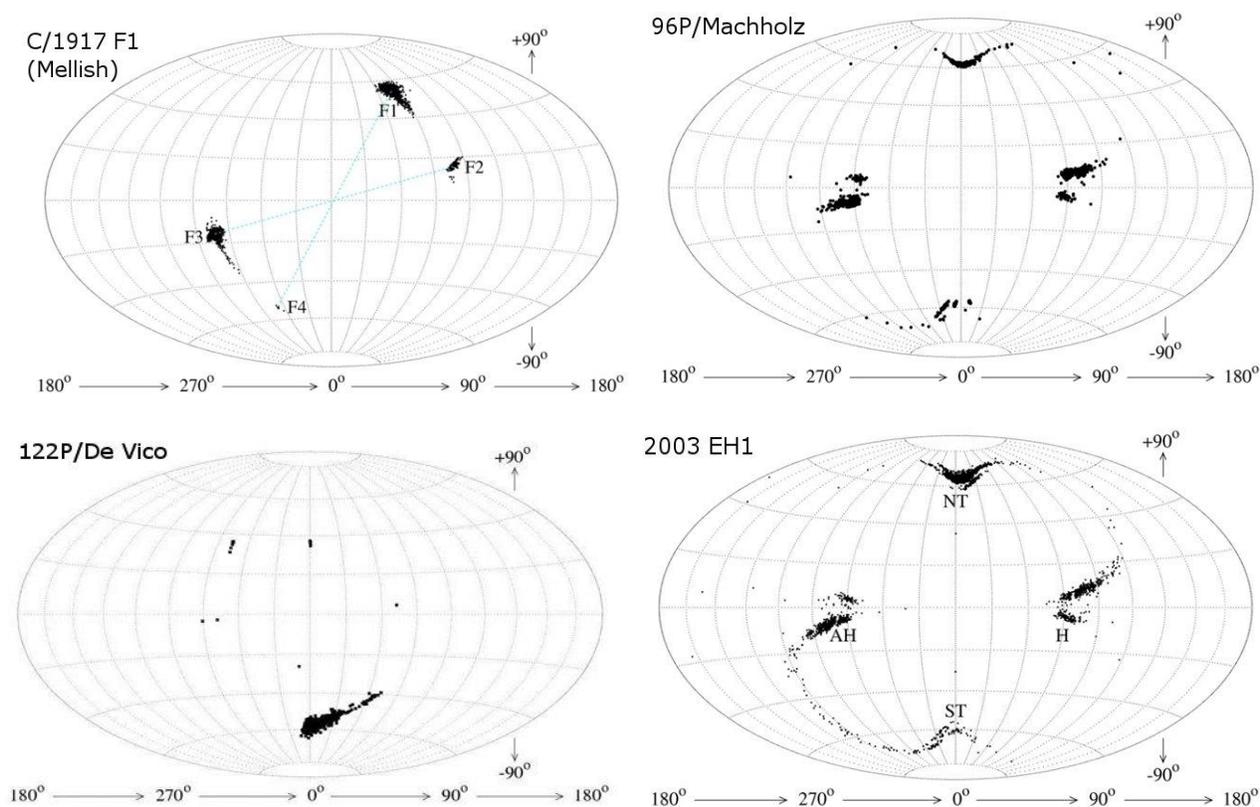

Figure 5- The radiants of showers are distributed on the sky symmetrically with respect to the Earth's apex. Some of them are symmetrical to the ecliptic (e.g. showers of 96P/Machholz and asteroid 2003 EH1, which create an ecliptic-toroidal structure). Some of them have no counterparts with a similar ecliptical longitude on the opposite hemisphere (C/1917 F1 Mellish). The ecliptical coordinate system is modified to place its origin at the apex of the Earth's motion.

## Acknowledgment

The work was also supported, by the VEGA – the Slovak Grant Agency for Science, grants No. 2/0031/14 and 1/0225/14 and by the Slovak Research and Development Agency under the contracts No. APVV-0517-12 and APVV-0158-11.

## References

Asher, D. J. (1999), The Leonid meteor storms of 1833 and 1966, MNRAS, 307, pp. 919-924

Asher, D. J. (2005), in Dynamics of Populations of Planetary Systems, eds. Z. Knežević, & A.Milani (Cambridge: Cambridge Univ. Press), Proc. IAU Coll., 197, 375

Asher, D. J. (2008), Meteor Outburst Profiles and Cometary Ejection Models, Earth Moon Planets, 102, 27- 33

Asher, D. J. & Emel'yanenko, V. V., (2002), The origin of the June Bootid outburst in 1998 and determination of cometary ejection velocities, MNRAS, 331, pp. 126-132

Babadzhanov, P. B., Williams I. P., & Kokhirova, G. I. (2008), Meteor showers associated with 2003EH1, MNRAS, 386, 2271

Babadzhanov, P. B., Williams I. P., & Kokhirova, G. I. (2013), Near-Earth asteroids among the Scorpiids meteoroid complex, A&A 556, A25, pp. 1-5

Beech, M. (2001), Comet 72P/Denning-Fujikawa: down but not necessarily out, MNRAS, 327, 1201

Brown, P., Jones, J. (1998), Simulation of the Formation and Evolution of the Perseid Meteoroid Stream, Icarus, Volume 133, Issue 1, pp. 36-68.

Brown P., Wong, D.K., Weryk, R.J., Wiegert, P. (2010), A meteoroid stream survey using the Canadian Meteor Orbit

Radar II: Identification of minor showers using a 3D wavelet transform, Icarus 207, 66–81

Campbell-Brown, M. D., & Brown, P. (2005), The Meteoroid Environment: Shower and Sporadic Meteors, LPI Contribution, 1280, 29

Jenniskens, P. (2004), 2003 EH$_1$ Is the Quadrantid Shower Parent Comet, AJ, 127, pp. 3018-3022

Jenniskens, P., Vaubaillon, J., (2010), "Minor Planet 2002 EX12 (=169P/NEAT) and the Alpha Capricornid Shower", AJ, Vol. 139, pp. 1822-1830

Jenniskens, P., (2006), Meteor Showers and Their Parent Comets, Cambridge Univ. Press, Cambridge

Jopek, T. J., & Williams, I. P., (2013), Stream and sporadic meteoroids associated with near-Earth objects, MNRAS, 430, 2377-2389

Kaňuchová, Z. and Neslušan, L. (2007), The parent bodies of the Quadrantid meteoroid stream, A&A 470, 1123

Lindblad, B. A., Neslusan, L., Porubcan, V., Svoren, J., (2003), IAU Meteor Database of photographic orbits version 2003, Earth Moon Planets, 93, 249

Lyytinen, E., & Jenniskens, P., (2003), Meteor outbursts from long-period comet dust trails, Icarus, 162, p. 443-452.

Neslušan, L. (1999), Comets 14P/Wolf and D/1892 T1 as parent bodies of a common, α-Capricornids related, meteor stream, A&A 351, 752

Neslušan, L., Kaňuchová, Z., and Tomko, D. (2013a), The meteor-shower complex of 96P/Machholz revisited, A&A 551, 14

Neslušan, L., Hajduková, M., jr., and Jakubík, M. (2013b), Meteor-shower complex of asteroid 2003 EH1 compared with that of comet 96P/Machholz, A&A 560, A47

Neslušan, L. and Hajduková, M., jr. (2014), The meteor-shower complex of comet C/1917 (Mellish), A&A, in press

Neslušan, L., Kaňuchová, Z., and Tomko, D. (2014), Ecliptic-toroidal structure of the meteor complex, in Proceedings of the Astronomical Conference held at A. M. University, Poznan, Poland, eds. T. J. Jopek, F. J. M. Rietmeijer, J. Watanabe, I.P. Williams, A.M. University Press, 235–242

Porubcan, V., Kornos, L. (2005), The Quadrantid meteor stream and 2003 EH1, Contrib. Astron. Obs. Skalnaté Pleso, 35, no. 1, p. 5-16

Rudawska, R., Vaubaillon, J. and Atreya P. (2012), Association of individual meteors with their parent bodies A&A 541, pp. A2, 1-5

Rudawska R., Vaubaillon, J., (2014), Don Quixote - a possible parent body of a meteor shower, In Gyssens M. and Roggemans P., editors, Proceedings of the International Meteor Conference 2014, Giron, France, submitted

Sekhar, A., & Asher, D. J. (2014a), Resonant behavior of comet Halley and the Orionid stream, Meteor. Planet. Sci., 49, pp. 52-62

Sekhar, A., & Asher, D. J. (2014b), Meteor showers on Earth from sungrazing comets, MNRAS, 437, L71-L75

SonotaCo, (2009), Journal of the IMO, 37, 55 (http://sonotaco.jp/doc/ SNM/)

Tomko, D. and Neslušan, L. (2012), Search for New Parent Bodies of Meteoroid Streams Among Comets. I. Showers of Comets 126P/1996 P1 and 161P/2004 V2 with Radiants on Southern Sky, EM&P 108, 123

Tomko, D. and Neslušan, L. (2014), Prediction of meteor shower of comet 161P/2004 V2, in Proceedings of the Astronomical Conference held at A. M. University, Poznan, Poland, eds. T. J. Jopek, F. J. M. Rietmeijer, J. Watanabe, I.P. Williams, A.M. University Press, 243–249

Tomko, D. (2014), Prediction of meteor shower associated with Comet 122P/de Vico, CAOSP 44, 33

Vaubaillon, J., Arlt, R., Shanov, S., Dubrovski, S., & Sato, M. (2005a), The 2004 June Bootid meteor shower, MNRAS, 362, 1463

Vaubaillon, J., Colas, F., & Jorda, L. (2005b), A new method to predict meteor showers. I. Description of the model, A&A, 439, pp. 751-760

Vaubaillon, J., & Jenniskens, P., (2007), Dust trail evolution applied to long-period comet C/1854 L1 (Klinkerfues) and the ε-Eridanids, Adv. Space Res., 39, pp. 612-615

Vereš, P., Kornoš, L., & Tóth, J., (2011), Meteor showers of comet C/1917 F1 Mellish, MNRAS, 412, 511

Wiegert, P. and Brown, P., (2004), The problem of linking minor meteor showers to their parent bodies: Initial considerations, Earth, Moon, and Planets 95, 19–25

Wiegert, P., & Brown P. (2005), The Quadrantid meteoroid complex, Icarus, Volume 179, Issue 1, p. 139-157

Wiegert, P. A., Brown, P. G., Vaubaillon, J., & Schijns, H., (2005), The tau Herculid meteor shower and Comet 73P/Schwassmann-Wachmann 3, MNRAS, 361, pp. 638-644

Wiegert, P., Vaubaillon, J., & Campbell-Brown, M.. (2009), A dynamical model of the sporadic meteoroid complex, Icarus, 201, pp. 295-310

Williams, I. P., & Wu, Z., (1994), The Current Perseid Meteor Shower, MNRAS 269, 524

Williams, I. P., Ryabova, G. O., Baturin, A. P., & Chernitsov, A. M., (2004), The parent of the Quadrantid meteoroid stream and asteroid 2003 EH1, MNRAS, 355, 1171